\documentstyle[preprint,aps]{revtex}
\begin{document}
\def\phik{\phi_{\bf k}}
\def\psik{\psi_{\bf k}}
\def\thetadot{\dot\theta}
\def\bfk{{\bf k}}
\def\gev{{\rm \,Ge\kern-0.125em V}}
\def\gtwid{\mathrel{\raise.3ex\hbox{$>$\kern-.75em\lower1ex\hbox{$\sim$}}}}
\def\ltwid{\mathrel{\raise.3ex\hbox{$<$\kern-.75em\lower1ex\hbox{$\sim$}}}}
\preprint{ACT-03/01, CTP-TAMU-11/01, hep-ph/0103348}
\draft
\title{Inflationary Baryogenesis}
\author{
RAGHAVAN RANGARAJAN$^{(a)}$
AND
D.V. NANOPOULOS$^{(b),(c),(d)}$}
\address{
$^{(a)}$ Theoretical Physics Division, Physical Research Laboratory\\
Navrangpura, Ahmedabad 380 009, India\\
$^{(b)}$ Department of Physics, 
Texas A\&M University\\
College Station, TX 77843-4242, USA\\
$^{(c)}$ Astroparticle Physics Group, Houston Advanced Research Center (HARC)\\
Mitchell Campus, The Woodlands, TX 77381, USA\\
$^{(d)}$ Chair of Theoretical Physics, Academy of Athens\\
Division of Natural Sciences, 28 Panepistimiou Avenue\\
Athens 10679, Greece}
\maketitle
\begin{abstract}

In this letter we explore the possibility of creating the baryon
asymmetry of the universe 
during inflation and reheating
due to the decay of a field associated with
the inflaton.
CP violation is attained by assuming
that this field is complex with a phase that varies as the inflaton
evolves.  We consider chaotic and natural inflation scenarios.  In the former
case, the complex decaying field is the inflaton itself and, in the latter case,
the phase of the complex field is the inflaton.
We calculate the asymmetry produced using the Bogolyubov
formalism that relates annihilation and creation operators at late time
to the annihilation and creation operators at early time.

\end{abstract}
\pacs{}

\narrowtext

\section{Introduction}

Explaining the origin of the matter-antimatter asymmetry of the universe is
an essential ingredient in our understanding of the history of the universe.
In this article, we study the possibility of creating the baryon
asymmetry of the universe by the production of particles
during inflation and reheating by the decay of a 
complex field related to the inflaton.
We consider the case where the complex decaying field is the inflaton itself
as well as the case where the phase of the complex field is the inflaton, as
in natural inflation.  The former case is similar to chaotic inflation but with
a complex inflaton.
By assigning baryon number to a scalar field present during
inflation and introducing a baryon number violating coupling between
this field and the inflaton
we find that there is a net baryon number asymmetry in the
produced particles.  

The notion that the inflaton plays a role in baryogenesis is not new.  
If the reheat temperature is above the mass of certain
heavy particles, such as GUT gauge and Higgs bosons,  then the latter are
thermally produced and their subsequent 
out-of-equilibrium decays create a baryon asymmetry.~\cite{napwein}  
The production of 
heavy GUT gauge and Higgs bosons or squarks by the direct decay of the
inflaton when the reheat temperature and/or the inflaton mass is
less than that of the heavy bosons has also been considered.  Once again,
the out-of-equilibrium decays or
annihilations of these
particles gives rise to the baryon asymmetry of the universe.
In all the above
scenarios 
\cite{dolgovlinde,dimhall,kolblinderiotto,chungkolbriotto}  
CP violation enters into the couplings of the heavier 
bosons to lighter
particles.  In our scenario the baryon asymmetry is produced in the direct 
decay of a field associated with the inflaton.  
Furthermore the CP violation must manifest itself
in the decay of this field.  In this respect it
is similar to the scenario mentioned in
Ref.~\cite{astw} and discussed in more detail in Ref.~\cite{kolbturnerb} in
which the baryon asymmetry is created by the direct decay of the inflaton.
However, unlike in Ref.~\cite{kolbturnerb}, in our scenario
CP violation is provided dynamically through the time dependent phase of an
evolving complex inflaton or of a complex field associated with the inflaton.  
We explicitly calculate the asymmetry in our scenario 
and compare it to the 
baryon asymmetry of the universe.
We follow the work of Ref.~\cite{papastparker79} in which the
asymmetry was calculated 
in the context
of a universe that contracts to a minimum size, bounces back and 
then expands.  The universe was static at both initial and late times.  The 
B-violating coupling was 
$\lambda R (\phi^*\Lambda\psi +\psi^*\Lambda^*\phi)$, where $R$ is the 
Ricci scalar, $\phi$ carried baryon number 
and $\Lambda$ was a complex function of time, which provided the necessary
CP violation.
In this work, we have adapted the formalism of Ref.~\cite{papastparker79}
to consider the asymmetry that might be created in a more realistic universe
that is initially inflating and then enters a reheating phase followed by
the standard evolution of the universe.  Furthermore we have given a more
realistic source of CP violation, namely, a time varying complex field.

Recently Ref.~\cite{fkot} appeared in which the authors discuss the generation
of the baryon asymmetry during preheating in a scenario similar to the one
discussed here.  We discuss later the differences and
similarities between our work and theirs.

As in Ref.~\cite{papastparker79}, we consider a lagrangian consisting of
two complex scalar fields $\phi$ and $\psi$.  $\phi$ and $\psi$
are assumed to carry
baryon number +1 and 0 respectively
and we assume a B-violating term
\begin{equation}
\lambda  (\eta^2 \phi^*\psi +\eta^{*2}\psi^*\phi),
\label{eq:bviol}
\end{equation}
where $\lambda$ is a dimensionless constant 
and $\eta$ is related to the inflaton field and is complex.
The baryon number of $\phi$ and $\psi$
particles is established by their interactions with other particles 
in the Standard Model.
\footnote{To ensure that the baryon asymmetry created is not erased by 
sphaleron processes, we assume that the interaction in Eq.~(\ref{eq:bviol})
also violates B-L.  This may be achieved, for example, if $\psi$ carries no
lepton number.}
The latter are not included in our lagrangian below as they do not enter into
our calculations.  
We assume that the initial velocity of the $\eta$ field and/or 
the shape of its
potential ensures that its phase varies as the inflaton
rolls down its potential.  Thus we have dynamic CP violation.

To obtain the asymmetry in our scenario we use 
the fact that the annihilation and creation
operators for the fields $\phi$ and $\psi$ are not the same during the
inflationary phase and 
at late times after reheating.
However, the annihilation and
creation operators 
at late times
can be written
as linear combinations of the annihilation and creation operators during
the inflationary phase using the Bogolyubov coefficients.  So,
\begin{eqnarray}
\tilde a^\phi_\bfk&=&A_k a^\phi_\bfk +B_{k}b^{\phi \dagger}_{-\bfk}
+A^\prime  a^\psi_\bfk +B^\prime_{k}b^{\psi \dagger}_{-\bfk}
\, ,\label{eq:Bog1}\\
\tilde b^{\phi\dagger}_\bfk&=&C_k a^\phi_{-\bfk} +D_{k}b^{\phi \dagger}_{\bfk}
+C^\prime  a^\psi_{-\bfk} +D^\prime_{k}b^{\psi \dagger}_{\bfk}
\, ,\label{eq:Bog2}
\end{eqnarray}
where $\tilde a^\phi_\bfk$ and $\tilde b^\phi_\bfk$ are operators 
at late times 
and $a^\phi_\bfk$ and $b^\phi_\bfk$ are operators at 
early times in the
inflationary phase.  Similar expressions exist for $\tilde a^\psi_\bfk$
and $\tilde b^\psi_\bfk$.  
In the Heisenberg picture, if we choose the
state 
to be the
initial vacuum state, 
then it will remain 
in that state during its subsequent
evolution.  One can then see that the number of $\phi$ particles and
antiparticles of
momentum $\bfk$, given by $\langle 0|\tilde
a_\bfk^{\phi\dagger} \tilde a^\phi_\bfk|0\rangle$ and 
$\langle0|\tilde b_\bfk^{\phi\dagger} \tilde b^\phi_\bfk
|0\rangle$ respectively,
are non-zero and proportional to $|B_k|^2+|B_k^\prime|^2$ and
$|C_k|^2+|C_k^\prime|^2$.
Furthermore, if $|B_k|^2+|B_k^\prime|^2\neq|C_k|^2+|C_k^\prime|^2$
then one obtains a baryon number asymmetry.  (If $\lambda=0$ in
Eq.~\ref{eq:bviol} then $B^\prime$ and $C^\prime$ are 0 and $|B|=|C|$, and
one gets no asymmetry.)  
 
Particle number, and correspondingly annihilation and creation operators,
are well defined only in adiabatic vacuum states.  However, the vacuum state
in the inflationary era, i.e., our {\it in} state, 
must be chosen judiciously to avoid infrared divergences.
This is discussed in Section III.

The framework of this article is as follows.  In Section II we present the
lagrangian density for the complex scalar fields $\phi$ and $\psi$
relevant to our calculation and obtain
their equations of motion.  We then write down the Fourier decomposition of
$\phi$ and $\psi$ during the inflationary phase and during reheating.
General expressions for the coefficients
$A-D^\prime$ in Eqs.~\ref{eq:Bog1} and ~\ref{eq:Bog2} have been derived
in Ref.~\cite{papastparker79}.  We shall present these results without
rederiving them and then present the general result for the baryon asymmetry of
the universe.  In Section III, we present the particular solutions for
our scenario of a universe that undergoes 
exponential inflation $(a\sim e^{Ht})$ 
followed by an inflaton-oscillation dominated phase $(a\sim t^{2/3})$.  
We then calculate the total baryon asymmetry for this scenario in the
context of chaotic and natural inflation.  
We conclude in the last section.
In the Appendix we discuss issues related to the regularisation of infrared
divergences and the necessity of an infrared cutoff to satisfy the 
conditions of perturbation theory.
\section{}

Consider a lagrangian density
\begin{eqnarray}
L&=&\sqrt{-g}[g^{\mu\nu}\partial_\mu\phi^*\partial_\nu\phi
+g^{\mu\nu}\partial_\mu\psi^*\partial_\nu\psi
-(m_\phi^2+\xi_\phi R)\phi^*\phi
-(m_\psi^2+\xi_\psi R)\psi^*\psi\\
&&+g^{\mu\nu}\partial_\mu\eta^*\partial_\nu\eta
-m^2\eta^*\eta -V(\eta)
-\lambda  (\eta^2 \phi^*\psi +\eta^{*2}\psi^*\phi)],
\label{eq:lagr}
\end{eqnarray}
where $m_{\phi,\psi}$ are the masses of the respective fields and 
$\xi_{\phi,\psi}$ are their couplings to the curvature.  We have assumed that
$\eta$ 
is minimally coupled.  $V(\eta)$ includes all interactions of $\eta$
other than the coupling to $\phi$ and $\psi$
already listed above.  Below we shall consider natural inflation and
chaotic inflation scenarios.  In the former case the inflaton will be
associated with the phase of $\eta$.  In the latter case we 
assume that
the complex $\eta$ field is the inflaton field.  (Thus we are really considering
an extension of chaotic inflation since the inflaton is now complex.)
We shall assume
a spatially flat Robertson-Walker metric.  The equations
of motion for the above fields are
\begin{eqnarray}
\ddot\phi+3\bigl({\dot a\over a}\bigr)\dot\phi-{1\over a^2}\nabla^2\phi
+(m_\phi^2+\xi_\phi R)\phi &+&\lambda \eta^2\psi=0,\label{eq:eqm1}\\
\ddot\psi+3\bigl({\dot a\over a}\bigr)\dot\psi-{1\over a^2}\nabla^2\psi
+(m_\psi^2+\xi_\psi R)\psi &+&\lambda \eta^{*2}\phi=0.\label{eq:eqm2}
\end{eqnarray}
We now write
\begin{equation}
\phi(x)=\sum_{\bf k} e^{i {\bf k\cdot x}}{1\over[La(t)]^{3/2}}
\phi_{\bf k}(t),\label{eq:phikk}
\end{equation}
with ${\bf k}={2\pi\over L}(n_x,n_y,n_z)$ and a similar expression for 
$\psi(x)$.
The equations of motion for the Fourier coefficients $\phik$ and $\psik$
are (note that $\phik$ and $\psik$ are operators)
\begin{equation}
\ddot\phik(t)+\Bigl({\bfk^2\over a^2}-{3\over4}{{\dot a}^2\over a^2}
-{3\over2}{\ddot a\over a} +m_\phi^2+\xi_\phi R \Bigr)\phik(t) 
 +\lambda R \Lambda\psik(t)=0,
\label{eq:eqmphik}
\end{equation}
\begin{equation}
\ddot\psik(t)+\Bigl({\bfk^2\over a^2}-{3\over4}{{\dot a}^2\over a^2}
-{3\over2}{\ddot a\over a} +m_\psi^2+\xi_\psi R \Bigr)\psik(t) 
 +\lambda R \Lambda^*\phik(t)=0,\label{eq:eqmpsik}
\end{equation}
$\phik$ and $\psik$ satisfy
\begin{equation}
[\phik(t),{\dot \phi_{\bf k^\prime}}^*(t)]=i\delta_{\bf k,
k^\prime},\label{eq:comm1}
\end{equation}
\begin{equation}
[\psik(t),{\dot \psi_{\bf k^\prime}}^*(t)]=i\delta_{\bf k,
k^\prime}.\label{eq:comm2}
\end{equation}

To use the results derived in Ref.~\cite{papastparker79} we
shall have to solve Eqs.~\ref{eq:eqmphik} and ~\ref{eq:eqmpsik}
for times
during inflation and during reheating.
To
simplify our calculations we shall assume that the fields $\phi$ and
$\psi$ are massless and minimally coupled, i.e.,
$m_{\phi,\psi}=0,\xi_{\phi,\psi}=0$.  (Typically, a spin zero particle will
obtain a mass of order $H$ during inflation, if $H$ is greater than its
bare mass\cite{mHinf}.  
However, this does not occur if the mass is protected by a
symmetry.)

To facilitate the use of perturbation theory and to be able to define
an {\it in} state we assume that 
the B-violating interaction switches on at some time $t_1$.  
Let $t_2$ be the time when
inflation ends, $t_3$ be the time when reheating ends
and $t_f$ be the final 
time at which we evaluate the baryon asymmetry.  
We assume that B-violation vanishes after $t_3$.
The annihilation and creation operators at $t_f$ can be expressed as
linear combinations of
the annihilation and creation operators at an early time $t_i$ before
$t_1$ in the inflationary era.  These relations have been derived perturbatively
to order $\lambda^2$
in Ref.~\cite{papastparker79} giving
\begin{eqnarray}
a_{f,\bfk}^\phi=&[\alpha_k^\phi(1+i\lambda^2 H_1^\phi)-i\lambda^2\beta_k^{\phi
*}H_3^\phi]a_{i,\bfk}^\phi
+[\beta^{\phi *}_k(1-i\lambda^2 H_4^\phi)+i\lambda^2\alpha_k^{\phi}H_2^\phi]  
b_{i,-\bfk}^{\phi \dagger}\cr
&-i\lambda[\alpha_k^{\phi} I_1-\beta_k^{\phi *} I_3]a_{i,\bfk}^\psi
-i\lambda[\alpha_k^{\phi} I_2-\beta_k^{\phi *} I_4]
b^{\psi\dagger}_{i,-\bfk}\, ,
\label{eq:aphi3a}\\
b_{f,\bfk}^{\phi\dagger}=&[\alpha_k^{\phi *}(1-i\lambda^2 H_4^\phi)+i\lambda^2
\beta_k^{\phi
}H_2^\phi]b_{i,\bfk}^{\phi\dagger}
+[\beta^{\phi }_k(1+i\lambda^2 H_1^\phi)-i\lambda^2\alpha_k^{\phi *}H_3^\phi]  
a_{i,-\bfk}^{\phi }\cr
&+i\lambda[\alpha_k^{\phi *} I_4-\beta_k^{\phi } I_2]b_{i,\bfk}^{\psi\dagger} 
+i\lambda[\alpha_k^{\phi *} I_3-\beta_k^{\phi } I_1]
a^{\psi}_{i,-\bfk}\, ,
\label{eq:bphi3a}\\
a_{f,\bfk}^\psi=&[\alpha_k^\phi(1+i\lambda^2 H_1^\psi)-i\lambda^2\beta_k^{\psi
*}H_3^\psi]a_{i,\bfk}^\psi
+[\beta^{\psi *}_k(1-i\lambda^2 H_4^\psi)+i\lambda^2\alpha_k^{\psi}H_2^\psi]  
b_{i,-\bfk}^{\psi \dagger}\cr
&-i\lambda[\alpha_k^{\psi} I_1^*-\beta_k^{\psi *} I_2^*]a_{i,\bfk}^\phi 
-i\lambda[\alpha_k^{\psi} I_3^*-\beta_k^{\psi *} I_4^*]
b^{\phi\dagger}_{i,-\bfk}\, ,
\label{eq:apsi3a}\\
b_{f,\bfk}^{\psi\dagger}=&[\alpha_k^{\psi *}(1-i\lambda^2 H_4^\psi)+i\lambda^2
\beta_k^{\psi
}H_2^\psi]b_{i,\bfk}^{\psi\dagger}
+[\beta^{\psi }_k(1+i\lambda^2 H_1^\psi)-i\lambda^2\alpha_k^{\psi *}H_3^\psi]  
a_{i,-\bfk}^{\psi }\cr
&+i\lambda[\alpha_k^{\psi *} I_4^*-\beta_k^{\psi } I_3^*]
b_{i,\bfk}^{\phi\dagger}
+i\lambda[\alpha_k^{\psi *} I_2^*-\beta_k^{\psi } I_1*]
a^{\phi}_{i,-\bfk}\, ,
\label{eq:bpsi3a}
\end{eqnarray}
where
\begin{eqnarray}
I_1=&\int^\infty_{-\infty} dt\, \eta^2(t) 
\chi^{\phi *}_k(t)
\chi^{\psi}_k(t)\, &,
\label{eq:I1}\\
I_2=&\int^\infty_{-\infty} dt\, \eta^2(t) 
\chi^{\phi *}_k(t)
\chi^{\psi *}_k(t)\, &,
\label{eq:I2}\\
I_3=&\int^\infty_{-\infty} dt\, \eta^2(t) 
\chi^{\phi }_k(t)
\chi^{\psi}_k(t)\, &,
\label{eq:I3}\\
I_4=&\int^\infty_{-\infty} dt\, \eta^2(t) 
\chi^{\phi }_k(t)
\chi^{\psi *}_k(t)\, &,
\label{eq:I4}
\end{eqnarray}
and
\begin{eqnarray}
H_1^\phi=&\int^\infty_{-\infty} dt\, \int^\infty_{-\infty} dt^\prime \,
\eta^2(t)\chi^{\phi *}_k(t) \cr
&\times
\triangle^\psi_k(t,t^\prime)\chi^{\phi}_k(t^\prime)\eta^{*2}(t^\prime)\,
&,
\label{eq:H1}\\
H_2^\phi=&\int^\infty_{-\infty} dt\, \int^\infty_{-\infty} dt^\prime \,
\eta^2(t)\chi^{\phi *}_k(t) \cr
&\times
\triangle^\psi_k(t,t^\prime)\chi^{\phi *}_k(t^\prime) \eta^{*2}(t^\prime)\,
&,
\label{eq:H2}\\
H_3^\phi=&\int^\infty_{-\infty} dt\, \int^\infty_{-\infty} dt^\prime \,
\eta^2(t)\chi^{\phi }_k(t)\cr
&\times
\triangle^\psi_k(t,t^\prime)\chi^{\phi}_k(t^\prime)\eta^{*2}(t^\prime)\,
&,
\label{eq:H3}\\
H_4^\phi=&\int^\infty_{-\infty} dt\, \int^\infty_{-\infty} dt^\prime \,
\eta^2(t)\chi^{\phi }_k(t) \cr
&\times
\triangle^\psi_k(t,t^\prime)\chi^{\phi *}_k(t^\prime)\eta^{*2}(t^\prime)\,
&,
\label{eq:H4}\\
\end{eqnarray}
and the $H_i^\psi$ are defined as $H_i^\phi$ with $\eta^2$ replaced by
$\eta^{*2}$ and $\chi^\phi_k$, $\chi^{\phi *}_k$ and $\triangle^\psi$
replaced by $\chi^\psi_k$, $\chi^{\psi *}_k$ and $\triangle^\phi$
respectively.
Above, $\chi^\phi_k$ and $\chi^\psi_k$ are complex functions (not
operators) that solve Eqs.~\ref{eq:eqmphik} and ~\ref{eq:eqmpsik},
with $\lambda$ set to 0, respectively.
$\triangle^\phi_k$ and
$\triangle^\psi_k$ are the retarded Green's functions for
Eqs.~\ref{eq:eqmphik} and ~\ref{eq:eqmpsik} respectively, i.e., they
satisfy Eqs.~\ref{eq:eqmphik} and ~\ref{eq:eqmpsik} with $\lambda$ set
to 0 and a delta function $\delta(t-t^\prime)$ on the r.h.s. of the
equations.
The subscripts on the coefficients $\alpha_k$ and $\beta_k$
and on the functions $\chi_k$  and $\triangle_k$ 
refer to $|\bfk|$. 
$\alpha_k$ and $\beta_k$ are complex and satisfy
\begin{equation}
|\alpha_k^\phi|^2-|\beta_k^\phi|^2=|\alpha_k^\psi|^2-|\beta_k^\psi|^2=1\,
.
\label{eq:Bog}
\end{equation}

We assume that the initial state at $t_i$ is the vacuum state. 
Then, in the Heisenberg picture, 
the number of $\phi$ particles of momentum $\bfk$ at $t_f$ is given by
\begin{eqnarray}
\langle N^\phi_\bfk(t_f)\rangle&=\langle in|a^{\phi\dagger}_{f,\bfk}
a^\phi_{f,\bfk}|in \rangle\cr
&=
|\beta^\phi_k|^2
+\lambda^2\Bigl[|\alpha^\phi_k|^2 |I_2|^2+|\beta^\phi_k|^2(|I_3|^2)+2Re
\bigl(\alpha^\phi_k\beta^\phi_k(i H^\phi_2-I_2 I_4^*)\bigr)\Bigr]
\label{eq:Nphik}
\end{eqnarray}
and
\begin{eqnarray}
\langle \bar N^\phi_\bfk(t_f)\rangle&=\langle 0,in|b^{\phi\dagger}_{f,\bfk}
b^\phi_{f,\bfk}|0,in \rangle\cr
&=|\beta^\phi_k|^2
+\lambda^2\Bigl[|\alpha^\phi_k|^2 |I_3|^2+|\beta^\phi_k|^2(|I_2|^2)+2Re
\bigl(\alpha^{\phi*}_k\beta^{\phi_k *}(i H^\phi_3+I_1^* I_3)\bigr)\Bigr]
\label{eq:Nbarphik}
\end{eqnarray}
where we have used
\begin{eqnarray}
2 Im H_1=|I_1|^2-|I_2|^2 \, ,
\label{eq:ImH1}\\ 
2 Im H_4=|I_3|^2-|I_4|^2 \,. 
\label{eq:ImH2} 
\end{eqnarray}
The baryon asymmetry for particles of momentum $\bfk$ at $t_f$ is thus
\begin{equation}
\triangle N^\phi_\bfk(t_f)=\langle N^\phi_\bfk(t_f)\rangle
-\langle \bar N^\phi_\bfk(t_f)\rangle=
\lambda^2(|I_2|^2-|I_3|^2)
\, ,
\label{eq:DeltaNphik}
\end{equation}
where we have used Eq.~\ref{eq:Bog} and 
\begin{equation}
H^\phi_2-H_3^{\phi *}=i I_1 I_3^*-i I_2 I_4^* \, .
\label{eq:H23}
\end{equation}
The reader is referred to Ref.~\cite{papastparker79} for a more detailed
derivation of the above results.  
\footnote
{Eqs.~(4.1c) and (4.9) of Ref.~\cite{papastparker79}
contain typographical errors.  The corrected equations are displayed here
in Eqs.~\ref{eq:apsi3a} and ~\ref{eq:H23} respectively.}
Note that the asymmetry does not
depend on $\alpha_k$ and $\beta_k$ implying that the asymmetry is
independent of the purely gravitational production of particles in the 
expanding universe.  ($\xi=0$ does not imply a conformally invariant
universe.  Therefore there is non-zero purely gravitational production
of particles in our scenario but it does
not contribute to the asymmetry.)

To obtain the net baryon number at $t_f$ we sum over all momentum modes
and take the continuum limit.  
Since we ultimately wish to express the baryon asymmetry as the baryon
number density to entropy density ratio, and the baryon number does not change
after $t_3$, we write
\begin{eqnarray}
n_B(t_3)&=\lim_{L\rightarrow\infty}
1/ ([L a(t_3)]^3)\sum_\bfk \langle \triangle
N^\phi_\bfk\rangle\cr
&=\lim_{L\rightarrow\infty}
1/ ([L a(t_3)]^3) \bigl({L\over 2\pi}\bigr)^3\int d^3k\langle \triangle
N^\phi_\bfk\rangle\cr
&=
1/  (2\pi^2[a(t_3)]^3) \int_0^\infty dk k^2 \langle \triangle
N^\phi_\bfk\rangle \, .
\label{eq:nB}
\end{eqnarray}

Working in the approximation that at $t_3$ the inflaton completely decays and
the universe instantaneously
reheats to a temperature $T_3$, the baryon asymmetry of the universe is
\begin{equation}
BAU\equiv n_B/s={\lambda^2\over(2\pi^2 [a(t_3)]^3)}
\int_0^\infty dk\,k^2
(|I_2|^2-|I_3|^2)/[(2\pi^2/45) g_* T_{3}^3)\,,
\label{eq:BAU}
\end{equation}
where we have assumed that there is no dilution of the baryon asymmetry
due to entropy production during the subsequent evolution of the universe.
Note that because the effective coupling is a complex function of time
the baryon asymmetry is obtained at $O(\lambda^2)$ and not $O(\lambda^4)$.

For standard reheating, 
$t_3\approx \Gamma^{-1}$,
where $\Gamma$ is the dominant perturbative decay rate of the inflaton. 
We take $\Gamma={g^2\over 8\pi} m_{inf}$, corresponding to the decay of the inflaton
of mass $m_{inf}$ to some light fermion-antifermion pair.
Furthermore, $T_3=\left({30\over\pi^2 g_*}\right)^{{1\over4}}
\rho(t_3)^{{1\over4}}
=0.6g_*^{-{1\over4}}(M_{Pl}\Gamma)^{1\over2}$\cite{kolbturner},
where $\rho$ here refers to the inflaton
energy density. 
\footnote{
The final temperature $T_4$ at the end of reheating is 
also a function of the
interactions of the inflaton decay products which we have ignored.
See Ref.~\cite{davidsonsarkar}
and references therein for a discussion of thermalisation
of the decay products.  
}
$\rho(t_3)\approx \rho(t_2)
[a(t_2)/a(t_3)]^3$.  
We assume that the reheat temperature is not high enough for GUT
B-violating interactions to be in equilibrium and wipe out the asymmetry
generated in our scenario.
On the other hand, we do not restrict ourselves
to reheat temperatures below $10^8\gev$ to avoid the gravitino problem
\cite{gravitino}
as we consider the possibility that the gravitino might be very light.

\section{}

To obtain the baryon asymmetry, we need to evaluate $I_2$ and $I_3$.
This requires obtaining $\chi^\phi_k$ and $\chi^\psi_k$.
We shall need to perform the integral for $I_2$ and $I_3$ only from
$t_1$ to $t_3$ 
as B-violation vanishes
earlier than $t_1$
and after $t_3$.  
Solutions of Eqs.~\ref{eq:eqmphik} 
and ~\ref{eq:eqmpsik} for $\lambda=0$ and $a(t)=\sigma t^c, \;
(c\neq1,-1/3)$ 
and $a(t)=\sigma e^{Ht}$
have been obtained in 
Ref.~\cite{fordparker77a}.  Using them we get
\footnote{
Note that the definition of the mode functions $\chi_k$
in Ref.~\cite{papastparker79}
differs from that in Ref.~\cite{fordparker77a} by a factor of $a^{3/2}$.}
\begin{eqnarray}
\chi^{\phi,\psi}_k=[ a(t)]^{3\over2}\Bigl[c_1 \bigl({-1\over 3
a(t)^3 H}\bigr)^{1/2}
H_{3\over2}^{(1)}\bigl({-k \over a(t)H}\bigr)
+c_2 \bigl({-1\over 3
a(t)^3 H}\bigr)^{1/2}
H_{3\over2}^{(2)}\bigl({-k \over a(t)H}\bigr)\Bigr] \quad
{\rm for } \; t_1<t<t_2\, ,
\label{eq:chiphipsiinf}\\
\chi^{\phi,\psi}_k=[ a(t)]^{3\over2}\Bigl[c_1^\prime \bigl({-t\over 
a(t)^3}\bigr)^{1/2}
H_{3\over2}^{(1)}\bigl({-3k t\over a(t)}\bigr)
+c_2^\prime \bigl({-t\over 
a(t)^3}\bigr)^{1/2}
H_{3\over2}^{(2)}\bigl({-3k t\over a(t)}\bigr)\Bigr] \quad
{\rm for } \; t_2<t<t_3\, .
\label{eq:chiphipsiosc}
\end{eqnarray}
$H_\nu(z)$ are Hankel functions.  The commutation relations Eqs.~\ref{eq:comm1}
and ~\ref{eq:comm2} imply that
\begin{equation}
\chi_k^{\phi,\psi}\dot\chi_k^{\phi,\psi *}
-\chi_k^{\phi,\psi *}\dot\chi_k^{\phi,\psi }=i
\label{eq:wronskchi}
\end{equation}
and therefore the constants $c_{1,2}$ and
$c_{1,2}^\prime$ satisfy
\begin{eqnarray}
|c_2|^2-|c_1|^2=3\pi/4\, ,
\label{eq:wronskinf}\\
|c_2^\prime|^2-|c_1^\prime|^2=-3\pi/4 \, .
\label{eq:wronskosc}
\end{eqnarray}

The constants $c_1$ and $c_2$ define an initial vacuum state in the inflationary
era.  
If one makes a choice of the de Sitter invariant vacuum state ($c_1=0$ and
$c_2=\sqrt {3\pi/4}$) as the {\it in} state
then it is well known that such a state suffers from an infrared divergence.
One option then is to choose the constants $c_1$ and $c_2$ appropriately
so
as to cancel the infrared divergences even though such states will no longer
be de Sitter invariant.  
In Ref.~\cite{fordparker77a}
it is suggested that one may choose the constants $c_1$ and $c_2$ as below
so as to cancel the infrared divergences:
\footnote{
Other mechanisms have also been suggested to avoid the infrared divergences.
For example, see Ref. \cite{dolgoveinhorn}.}
\begin{eqnarray}
c_1=(k t_2/a(t_2))^{-p}\, ,
\label{eq:c1}\\
c_2=((k t_2/a(t_2))^{-2p}+{3\pi\over 4})^{1/2}\, ,
\label{eq:c2}
\end{eqnarray}
with $p>0$.  (We point out in the Appendix that there is also an upper limit
on $p$ that was not mentioned in Ref.~\cite{fordparker77a}.)
As we discuss in the Appendix, the nature of the infrared divergences
is slightly different in our case. Though the above choice 
for the constants $c_{1,2}$ with appropriately chosen values of $p$
would make
the final integral over $k$ infrared finite, the integrands for the intermediate
integrals over $t$ become very large for small values of $k$, irrespective of
the value of $p$.  This
leads to a problem with perturbation
theory since the latter requires that
$\lambda^2 |I_{2,3}|^2$ should be less than 1.  
\footnote{
We thank D. Lyth for pointing this out to us.}
This is discussed in more detail in the Appendix.
Therefore we are forced to introduce
a low momentum cutoff $k_L$ to justify our use of perturbation 
theory.  We choose $k_L$ such that $k_L/a_2=1/t_2$.
Since the low momentum cutoff automatically regulates the integral over $k$,
we then choose $c_1=0$ and $c_2=\sqrt{3\pi/4}$.
Continuity conditions
for $\phi(x)$, $\dot\phi(x)$ and $a(t)$ at $t_2$ imply that
$\chi(t)$ and $d/dt(\chi(t)/[a(t)]^{3/2})$ are continuous at $t_2$,
and these boundary conditions then give us $c_1^\prime$ and $c_2^\prime$. 
We have verified that the values of $c_1^\prime$ and $c_2^\prime$
obtained from the continuity conditions satisfy Eq.~\ref{eq:wronskosc}.

At this stage we need to specify $\eta(t)$.  If we write $\eta(t)$ as
${1\over\sqrt2}\sigma(t)e^{i\theta(t)}$ 
(where $\sigma(t)$ is real), it is the time varying
phase of $\eta$ that provides the CP violation necessary for creating a net
baryon asymmetry.  

{\bf Chaotic Inflation}

We first consider the case of chaotic inflation in which the $\eta$ field
represents a complex inflaton field.
In the absence of any potential for $\theta$, the equation of motion for
$\theta$ is
\begin{equation}
\sigma^2\ddot\theta + 3H\sigma^2\dot\theta+2\dot\sigma\sigma\dot\theta=0\,.
\label{eq:theom}
\end{equation}
In a more realistic model $V(\eta)$ will imply a potential for
$\theta$.  The equation of motion for $\sigma$ is
\begin{equation}
\ddot\sigma +3H\dot\sigma +m^2\sigma-\dot\theta^2\sigma=0\,.
\label{eq:sigeom}\end{equation}

We assume that $\theta(t)$ evolves starting from
an initial value of 0 at $t=t_i$ and an initial velocity $\dot\theta_i$.
We choose $\dot\theta_i$ consistent
with a universe dominated by the potential energy of $\sigma$.  Therefore
we take $\dot\theta_i=m/2$.  
During inflation, $\dot\sigma^2\ll m^2\sigma^2$ and $\sigma\sim M_{Pl}$.
Hence $\dot\sigma/\sigma\ll H$ and we ignore the last term in the equation
of motion for $\theta$.  
Then
\begin{equation}
\theta(t)={\dot\theta_i\over 3H} (1-e^{-3H(t-t_i)})\quad t_i\le t\le t_2
\label{eq:thetainfl}\end{equation}
We take
$H$ to be constant during inflation and corresponding to the
initial energy density of the universe with $\sigma(t_i)=3M_{Pl}$.  From 
above, one can see that $\thetadot$ is much less than $m$ 
for most of the inflationary era and so we ignore the last term of
Eq.~\ref{eq:sigeom} for this era.  Invoking the slow roll approximation one
may also ignore $\ddot\sigma$ during inflation.

During reheating the $\sigma$ and the $\theta$ fields are coupled and 
we can not ignore the last terms of their respective equations of
motion.  However, to obtain $\theta(t)$ it is simpler to first rewrite
$\eta$ as ${1\over\sqrt2}(\kappa_1 + i\kappa_2)$.  Then the problem reduces
to one of two uncoupled damped harmonic oscillators with solutions,
$\kappa_1=(A_1/t) cos(mt+\alpha)$ and 
$\kappa_2=(A_2/t) cos(mt+\beta)$.  Here we have assumed $H=2/(3t)$ during
reheating.  $\theta(t)$ is then $tan^{-1}(\kappa_2/\kappa_1)$.
The constants $A_1$, $A_2$, $\alpha$ and
$\beta$ are determined by the values of $\kappa_{1,2}$ 
and their time derivatives
at $t_2$ which can be obtained from the values of $\theta$ and $\sigma$ and
their time derivatives at $t_2$.  
We take $t_2\approx2/m$, 
as the inflaton starts oscillating when $3H\approx m$.
$\sigma(t_2)\approx M_{Pl}/6$.  

Eq.~\ref{eq:thetainfl} implies that $\theta$ becomes nearly constant within
a few e-foldings after $t_i$. 
If the B-violating interaction switches on subsequent to this
then $\theta$ is approximately constant between
$t_1$ and $t_2$.  Furthermore, since $\dot\theta(t_2)$ is practically zero,
there is practically no rotational motion during reheating
in the absence of any potential for $\theta$. So during reheating
$\theta$
takes values of $\theta_2$ and $\theta_2+\pi$ during different phases
of the oscillation of $\sigma$, 
where $\theta_2$ is the value at $t_2$.  
($\theta$ changes discontinuously at the bottom
of the potential where $\sigma$ is 0.)  Since the relevant 
phase in $I_2$ and $I_3$ is $2\theta$ the above implies that the CP phase is
practically the same for the interval $t_1$ to $t_3$ and hence one should
expect very little asymmetry.

{\bf Natural Inflation}

We now consider natural inflation, in which case $\sigma(t)=f$
where $f$ is the scale of spontaneous symmetry breaking in the natural
inflation scenario.
In the presence of an
explicit symmetry breaking term that gives mass $m_\theta$ to the inflaton
$\theta$ the equation of motion for $\theta$ is
\begin{equation}
f^2\ddot\theta + 3H f^2\dot\theta+m_\theta^2f^2\theta=0\,.
\label{eq:theomnat}
\end{equation}
We assume that $\theta$ is constant during inflation between $t_1$ and
$t_2$ and is of $O(1)$.  Our results are insensitive to $t_1$ for $t_1$
earlier than about 10 e-foldings before the end of inflation.
At $t_2\approx 2/m_\theta$ when 
$3H\approx m$ the $\theta$ fields starts oscillating in its potential.
Between $t_2$ and $t_3$ $\theta$ evolves as
\begin{equation}
\theta(t)=\theta(t_2)  {t_2\over t} cos[m_\theta(t-t_2)] \,.
\label{eq:theeomnatreh}\end{equation}

Now we obtain the asymmetry numerically.
Smaller the value of $g$,
longer is the period of reheating contributing to the asymmetry.  But
for $g\le 10^{-3}$, $I_2$ and $I_3$ become independent of $g$ and then the
$g$ dependence in $BAU$ enters through $a(t_3)$ and the reheat temperature
$T_3$. For $g\le 10^{-3}$, we get
\begin{equation}
BAU=\lambda^2 g (2\times 10^{10})\, .
\end{equation}
As we have mentioned before, perturbation theory requires that 
$\lambda |I_{2,3}|^2$ must be less than 1.  This translates into an upper
bound on $\lambda$ of $10^{-11}$.  Then even for $g=10^{-3}$
we get insufficient asymmetry.  Other values of $g$ give even less asymmetry.

\section{Conclusion}

In conclusion, we have discussed a mechanism for
creating a baryon asymmetry during inflation and reheating.
While the scenario illustrated above does not create sufficient asymmetry,
it may be easily modified to accommodate a potential for $\theta$ which
can give rise to a much larger asymmetry.  A possible potential for
$\theta$ for the chaotic inflation scenario
is $W(\theta)=m_\theta^2\sigma^2(1-\cos\theta)$, which is equivalent
to tilting the inflaton potential. 
Unlike in the analogous axion and natural inflation
models, here  both $\sigma$ and $\theta$ would be
varying with time. Hence such a potential may allow 
for chaotic orbits
and so would have to be studied with care.

We point out here that we include both the inflationary
phase and the reheating phase in our calculation.  Contributions during
both phases do get mixed up in the evaluation of the asymmetry because of
the presence of $|I_2|^2$ and $|I_3|^2$, where the time integrals in
$I_2$ and $I_3$ 
include both the inflationary and the reheating eras.  In fact we find in
the natural inflation case that though the phase is taken to be constant
during the inflationary era, the net baryon asymmetry for a fixed
value of $\lambda$ decreases
if we do not include the inflationary era in the integrals $I_2$ and $I_3$.
\footnote{The values of $|I_{2,3}|^2$ also decrease and so, in principle,
one may obtain a slightly larger asymmetry by choosing a larger value of $\lambda$
which is still consistent with the use of perturbation theory.}
This indicates that one should not ignore the inflationary era
when calculating the asymmetry. 

While we were writing up this paper Ref.~\cite{fkot} appeared.  In this paper
the authors discuss the generation of baryon asymmetry during reheating
in a scenario similar to ours.  The two calculations have some differences
however.  Our calculation is carried out in curved spacetime while the
mode functions in Ref.~\cite{fkot} are obtained in Minkowski space.  We 
consider standard reheating while they consider the more complicated 
preheating scenario.  In both calculations the source of CP violation is
a time varying phase.  The authors of Ref.~\cite{fkot} suggest that
the CP violating potential for the baryonic fields
may be induced by their direct coupling to the inflaton or through loop effects 
involving the baryonic fields and other fields and then presume a form for
the phase.  We provide a
specific scenario in which the inflaton, or a field related to the inflaton, 
which is coupled to the baryonic
fields, is complex and its time varying phase dynamically provides
CP violation.  
Involving the inflaton and its phase seems to us to be a simple
and a 
very natural approach to obtain a time dependent phase.  
Our calculation includes both the inflationary and the 
reheating eras which, as we have pointed out, seems to be appropriate
for our case.

\acknowledgements

R. R. would like to thank Sacha Davidson, 
A. D. Dolgov, Salman Habib, Arul Lakshminarayan, David Lyth, Leonard Parker 
and U. B. Sathuvalli for useful discussions and comments.
R. R. would also like to thank the University of Lancaster for their 
hospitality.
The work of D. V. N. was partially supported by DOE grant DE-FG03-95-ER-40917.

\section*{Appendix}

In Ref.~\cite{fordparker77a} infrared divergence in $\langle\phi(x)\phi(x^\prime)\rangle$
for Robertson-Walker universes with power law expansion and exponential expansion
is discussed.  (In this Appendix, $\phi$ is a generic scalar field.)
Now 
\begin{equation}
\langle \phi(x)\phi(x^\prime)\rangle\sim \int d^3k e^{i\bf{k}\cdot(\bf{x}-\bf{x^\prime})}
\chi_k^{FP}(\tau) \chi_k^{FP*}(\tau^\prime) \, ,
\end{equation}
where $\chi_k^{FP}(\tau)$ are mode functions as defined in Ref.~\cite{fordparker77a} (they are $a^{3/2}$ times the mode functions defined in this paper)
and $\tau=\int^t a^{-3}(t^{\prime\prime}) dt^{\prime\prime}$.  If one
argues that $H_{3\over2}^{(1)}(z)\sim H_{3\over2}^{(2)}(z)\sim iY_1(z)\sim z^{-3/2}$ as $z\rightarrow 0$ the integrand in the above integral $\sim |c_1-c_2|^2
k^{-3}$, which implies a divergent 2-point function unless one chooses $c_1(k)$ and $c_2(k)$ such that $|c_1-c_2|\rightarrow 0$ fast enough as $k\rightarrow 0$.
Hence the choice given in Eqs.~(\ref{eq:c1}) and (\ref{eq:c2}) 
is suggested in Ref.~\cite{fordparker77a} which
eliminates the infrared divergences if $p>0$.  We would like to point out
here that there is also an upper bound on $p$.
Unless $p$ is less than 3, 
the term proportional to $J_1^2$ in the 2-point
function, namely, $\sim |c_1+c_2|^2 J_1^2\sim k^{-2p+3}$ will also diverge
at low values of $k$.

Unlike in Ref.~\cite{fordparker77a} 
the divergent quantities for us are
$\sim \int d^3k  \chi_k^{\phi *} \chi_k^{\psi *}$
and
$\sim \int d^3k  \chi_k^{\phi } \chi_k^{\psi }$ and the form of the divergences
is slightly different.  
Adapting the arguments of Ref.~\cite{fordparker77a} 
we would choose ${3\over4}<p<{9\over4}$ for
the constants $c_1$ and $c_2$ given in Eqs.~(\ref{eq:c1}) and (\ref{eq:c2})
to avoid infrared divergences in the integral
over momentum 

In our case we also have intermediate integrals $I_2$ and $I_3$ over time
and this leads to an additional problem. 
The $k$ dependence in the integrand of, for example, $I_3$,  
is contained in $-(c_1-c_2)^2 Y_1^2(z) + (c_1+c_2)^2 J_1^2(z) +2i
(c_1^2-c_2^2) J_1(z)Y_1(z)$, where $z$ is the relevant argument
of the Bessel functions depending on whether the universe is in the
inflationary or the reheating era.  The $k$ dependence for $I_2$ is
similar.
When $k$ is small the argument of the Bessel functions 
becomes small.  For low $k$ values the first term goes as
$k^{2p-3}$, the second as $k^{3-2p}$ and the third is $k$
independent.
Now perturbation theory requires that $\lambda^2 |I_{2,3}|^2$ be less than 1.  
However, since $2p-3$ and $3-2p$ cannot both be greater than 0,
$I_2$ and $I_3$ will become very large at low $k$ values.  
Thus,
without a low momentum cutoff, perturbation theory breaks down at some
point for any finite value of $\lambda$.  
We emphasise again that this is an issue related to the
validity of perturbation theory
and not to the infrared divergence of  the integral over momentum.  The latter
can be regulated by the choice of 
constants $c_1$ and $c_2$ mentioned above,
irrespective of whether or not
perturbation theory is valid.


\begin{references}
\bibitem{napwein} D. V. Nanopoulos and S. Weinberg, 
Phys. Rev. D {\bf 20} (1979) 2484.

\bibitem{dolgovlinde} A. D. Dolgov and A. D. Linde, Phys. Lett. {\bf B116}
(1982) 329. 

\bibitem{dimhall} S. Dimopoulos and L. Hall, Phys. Lett. {\bf B196} (1987) 135. 

\bibitem{kolblinderiotto} E. W. Kolb, A. Linde and A. Riotto, Phys. Rev. Lett.
{\bf 77} (1996) 4290.

\bibitem{chungkolbriotto} D. Chung, E. W. Kolb and A. Riotto,
Phys. Rev. D {\bf 60} (1999) 063504.

\bibitem{astw} A. Albrecht, P. J. Steinhardt, M. S. Turner and F. Wilczek,
Phys. Rev. Lett. {\bf 48} (1982) 1437.

\bibitem{kolbturnerb} 
E. W. Kolb and M. S. Turner, ``The Early Universe",
(Addison-Wesley, Redwood City, California, 1990), 
see Secn. 6.7. 

\bibitem{papastparker79} N. J. Papastamatiou and L. Parker, 
Phys.\ Rev.\ D {\bf 19} (1979) 2283.  

\bibitem{fkot} K. Funakubo, A. Kakuto, S. Otsuki and F. Toyoda, hep-ph/0010266,
(2000).

\bibitem{mHinf} M. Dine, W. Fischler and D. Nemeschansky,
        Phys. Lett. {\bf B136} (1984) 169;
O. Bertolami and G. G. Ross,
        Phys. Lett. {\bf B183} (1987) 163;
E. J. Copeland, A. R. Liddle, D. H. Lyth, E. D. Stewart
        and D. Wands, Phys.\ Rev.\ D {\bf 49} (1994) 6410.

\bibitem{kolbturner} 
See Eq. 8.34 of Ref.~\cite{kolbturnerb}.

\bibitem{davidsonsarkar} S. Davidson and S. Sarkar, hep-ph/0009078,
(2000).

\bibitem{gravitino}
J. Ellis, A. D. Linde, and D. V. Nanopoulos, Phys. Lett. {\bf B118} (1982) 59;
M. Yu. Khlopov and A. D. Linde,  Phys. Lett. {\bf B138} (1984) 265;
J. Ellis, J. E. Kim and D. V. Nanopoulos, Phys. Lett. {\bf B145}
(1984) 181;
J. Ellis, D. V. Nanopoulos and S. Sarkar, Nucl. Phys. {\bf B259} (1985) 175.

\bibitem{fordparker77a} L. H. Ford and L. Parker, 
Phys.\ Rev.\ D {\bf 16} (1977) 245.

\bibitem{dolgoveinhorn} A. D. Dolgov, M. B. Einhorn and V. I. Zakharov,
Acta Phys. Polon. {\bf B26} (1995) 65.

\end{references}
\end{document}